# The Extremely Large Telescope Interferometer

Francisco Prada*
*Instituto de Astrofísica de Andalucía* CSIC, *Granada, Spain*
*Instituto de Astrofísica de Canarias, Tenerife, Spain*

Enrique Pérez
*Instituto de Astrofísica de Andalucía* CSIC, *Granada, Spain*

Sergio Fernández-Acosta
*Grantecan, La Palma, Spain*

Km Nitu Rai
*Aryabhatta Research Institute of Observational Sciences* (ARIES), *Nainital, India*

and

Joel Sánchez-Bermúdez
*Instituto de Astronomía, Universidad Nacional Autónoma de México, México*

* Contact: f.prada@csic.es

**Executive Summary**

The ELTI concept capitalizes on recent breakthroughs in large-format SPAD (Single-Photon Avalanche Diode) imaging sensors, combining them with the unprecedented collecting area and segmented architecture of the ELT to deliver a fundamentally new observational capability for the 2040s: visible-light intensity interferometry with ELT-scale angular resolution and quantum-limited temporal sampling. By uniting three transformative innovations - segmented sub-pupil beam combination, megapixel SPAD arrays with picosecond time resolution, and high-dispersion spectroscopy - ELTI opens an entirely new region of observational parameter space in angular resolution, spectral bandwidth, and temporal precision. This capability enables true milliarcsecond imaging of stellar surfaces, direct probes of extreme-gravity and strong electromagnetic-field environments around compact objects, precision studies of accretion physics, and even the potential detection of Earth-sized exoplanets through minute visibility modulations. ELTI thus positions the ELT as the world-leading facility for ultra–high-resolution optical astronomy in the coming decades.

# 1. The Extremely Large Telescope Interferometer (ELTI) at optical wavelengths

## 1.1. The ELTI configuration

The ELT primary mirror (M1) segments are grouped into sub-pupils to balance the trade-off between increasing the number of sub-pupils – which improves the uv-plane coverage – and increasing the collecting area of each sub-pupil, which enhances the signal-to-noise ratio. Figure 1 shows a proposed configuration in which 33 sub-pupils of 7.25 m each – arranged using 19 mirror segments – are formed using 627 of the 798 M1 segments (79% of the total). All sub-pupils are constructed with the same number of segments and arranged symmetrically to simplify post-processing. The segments belonging to each sub-pupil are shown in the same color, while unused segments appear in white. The black spots mark the geometric center of each sub-pupil.

In this configuration, the ELT Interferometer with 33 sub-pupils yields 528 independent baselines, computed using the relation *N(N-1)/2*. The maximum baseline - defined as the greatest separation between any pair of sub-pupil centers – is 33.5 m.

## 1.2. ELTI interferometry

For the sake of simplicity, we consider here the case of intensity interferometry in a single wavelength channel. The 19 segments forming each sub-pupil will be configured to produce a single image on the SPAD detector. This results in the detector being illuminated by 33 images with the same PSF arranged in this predefined pattern. This pattern will remain fixed throughout the observation. Figure 1 right shows the image on the SPAD camera corresponding to the ELTI configuration described above. In this example, we are adopting a SPAD detector with 2,000 × 2,000 pixels, each with a size of 20 μm. The separation between sub-pupil images ensures 90% of the flux for a seeing of 0.5 arcsec image at 4500 Å.

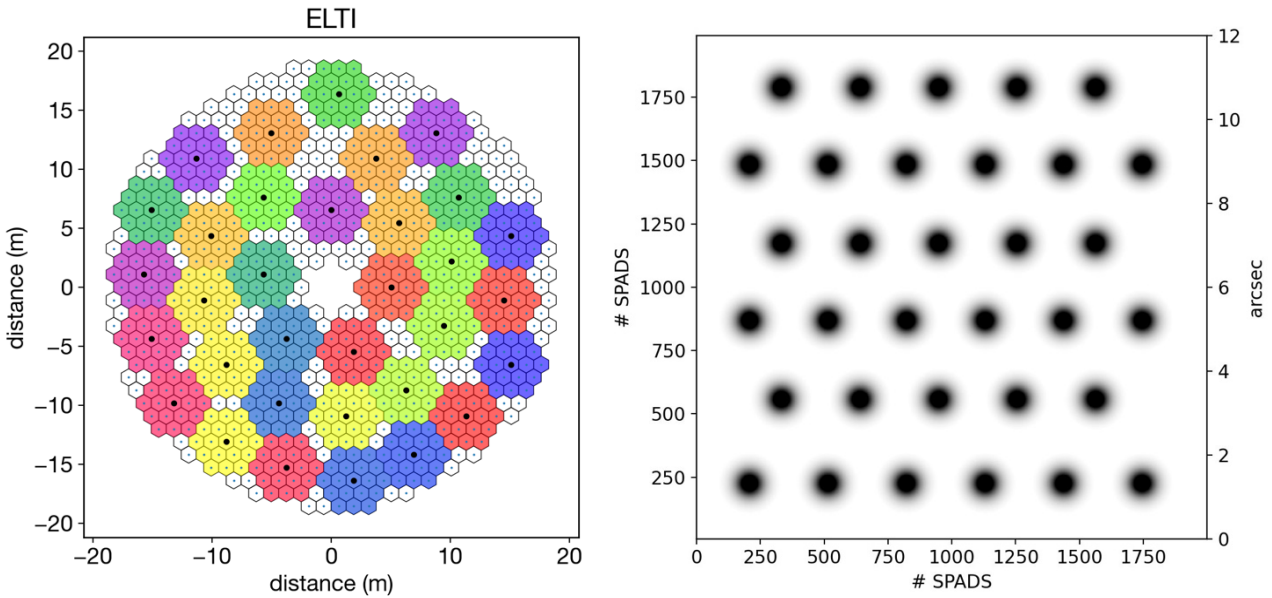

Figure 1. Left: Configuration of ELT M1 segments into 33 sub-pupils of 7.25 m each (colored). With this layout, 79% of the segmented mirrors are used (627 out of 798). The black spots indicate the center of each sub-pupil. Right: Image of the 33 sub-pupils on the SPAD detector; each circle encompasses 90% of the flux for each PSF of FWHM 0.5 arcsec at 4000 Å.

Thus, with this arrangement, we effectively perform visible-light interferometric observations equivalent to using 33 telescopes, each with a diameter of 7.25 m. Using the center position of each sub-aperture (black spots in Figure 1), we will have 528 baselines with lengths smaller than 33.5 m; as shown in the left panel of Figure 2. The right panel shows the simulation of the visibility for a uniformly illuminated disk with a diameter of 2.75 mas at 4500 Å.

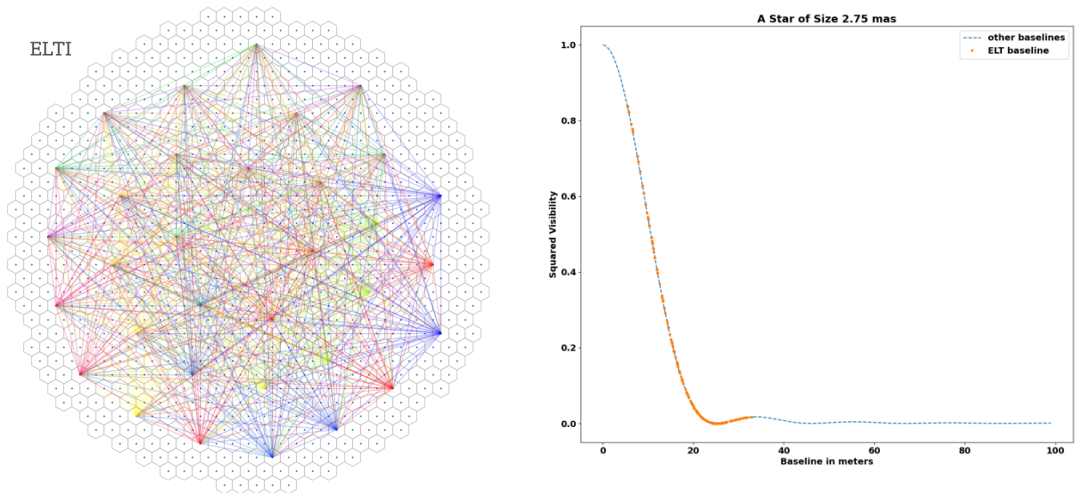

Figure 2. Left panel: ELTI "chakras" diagram illustrating the uv-plane coverage sampled by the 528 baselines of the 33 ELTI sub-pupils. The right panel shows the simulation of the visibility for a uniformly illuminated disk with a diameter of 2.75 mas at 4500 Å.

The signal-to-noise ratio ($S/N$) of the correlation between any two sub-apertures is given in Zampieri et al.:

$$S/N = N\,(\lambda/c)\,(\lambda/\Delta\lambda)\,\alpha\,\gamma^2\,\sqrt{T_{expo}/(2\,dt)}$$

Where $N$ denotes the photon rate incident on a single ELT sub-aperture. For a filter of central wavelength $\lambda$ = 4500 Å, spectral width $\Delta\lambda$ = 0.1 Å, and with c the speed of light, we adopt an overall instrumental efficiency $\alpha$ = 0.5. The quantity $\gamma^2$ represents the squared modulus of the complex degree of coherence at zero optical delay; for a fully coherent signal we normalize $\gamma^2$ = 1. The time sampling interval is dt = 10 ps, corresponding to the temporal resolution of the SPAD detector, and $T_{expo}$ is the integration time in seconds. Considering the 528 independent sub-aperture pairs available in the ELTI configuration, we estimate that achieving a signal-to-noise ratio of S/N = 5 for a V = 6 mag star requires an integration time of order 100 seconds at 4500 Å.

To extend ELT intensity-interferometry observations toward fainter magnitudes and to extract spectrally resolved visibility information, we propose to couple the SPAD detector to a high-resolution spectrograph. This instrument will disperse the incoming light into approximately 50,000 spectral channels (each with $\Delta\lambda \approx$ 0.1 Å) across the 3500–9000 Å range. Because the total S/N in intensity interferometry scales as the square root of the number of statistically independent spectral channels, such multi-wavelength operation provides a multiplicative gain in sensitivity. Under these conditions, ELTI is expected to reach targets as faint as g-SDSS ≈ 14-mag within 1 hour of integration.

## 2. ELTI instrumentation

The Extremely Large Telescope Interferometer (ELTI) builds upon a new generation of visible-wavelength photon-counting sensors and dedicated beam-combination optics to perform optical intensity interferometry across the 33 sub-pupils of the ELT primary mirror described above. The key enabling technology is the rapid advancement of large-format SPAD (Single-Photon Avalanche Diode) detectors, capable of registering individual photons with temporal resolutions of only a few tens of picoseconds, extremely low dark-count rates, and high photon-detection probabilities across the optical band. These characteristics make SPAD arrays uniquely suited for measuring the rapid intensity fluctuations required for optical intensity interferometry at visible wavelengths.

In recent years, SPAD technology has progressed remarkably. Modern devices now achieve picosecond timing resolution, low dark current, and high quantum efficiency, enabling applications far beyond the capabilities of traditional CCD, CMOS, or infrared detectors. Until recently, astronomical uses were limited mostly to single-pixel SPADs, which prevented spatially resolved imaging or spectroscopic applications. The emergence of two-dimensional SPAD arrays is transforming this landscape by enabling time-resolved imaging and spectroscopy of fast astronomical phenomena, as well as intensity interferometry in the visible. The first astronomical results obtained with a 64×64-pixel SPAD sensor developed by the Instituto de Microelectrónica de Sevilla (CSIC-US) demonstrate this potential; see Prada et al. for details on the detector technology and its first on-sky validation.

ELTI will directly benefit from ongoing Spanish investments in SPAD technology, in particular the large-format megapixel SPAD detector currently under competitive procurement through a CDTI Pre-Commercial Public Procurement (CPP) call, aimed at astronomy applications[1], and in particular for visible intensity interferometry at the GTC, adopting the same concept proposed here for the ELT, and for the La Palma Quantum Interferometer[2].

The analysis of ELTI's massive, multidimensional datasets will require significant computational power and sophisticated algorithms, with GPU-accelerated AI methods expected to further enhance processing speed and scientific return.

## 3. Acknowledgements

We want to acknowledge the pioneering vision of Prof. Cesare Barbieri and his collaborators, whose early work on quantum optical astronomy within the OWL project provided a forward-looking conceptual foundation that continues to inspire present-day efforts. We are grateful to Jörg-Uwe Pott (MPIA Heidelberg) for advice about the ELT, to Guillaume Bourdarot, Frank Eisenhauer, Reinhard Genzel at MPE Garching, and Robert Content for fruitful discussions. KNR acknowledges support for her stay at the IAA-CSIC during the writing of this contribution, through project AST22_00001_27, funded by the NextGenerationEU Recovery and Resilience Plan, the Andalusian Regional Government, and CSIC. FP and EP acknowledge support from the Spanish PERTE-Chip grant PDC2023-145909-I00 for SPAD development.

---

[1] https://www.cdti.es/licitacion-detectores-spads
[2] https://lapalmaqi.es/